\documentstyle[12pt]{article}
\newlength{\dinwidth}
\newlength{\dinmargin}
\begin{document}

\begin{titlepage}
\def\thefootnote{\fnsymbol{footnote}}
\baselineskip18pt
\thispagestyle{empty}
\begin{flushright}
\begin{tabular}{l}
FTUAM-92/08\\\vspace*{24pt}
June, 1992\\
\end{tabular}
\end{flushright}

\vspace*{1.5cm}

\begin{center}
             {\Large {\bf NUMERICAL STUDY OF YANG-MILLS  \\
                CLASSICAL SOLUTIONS ON THE TWISTED   \\
                                TORUS}}
\end{center}

\vskip72pt\centerline{\bf M. Garc\'{\i}a P\'erez\footnote{Bitnet
address: MARGA@EMDUAM11} and A. Gonz\'alez-Arroyo\footnote{Bitnet
address: ARROYO@EMDUAM11}}

\vskip12pt
\centerline{{\it Departamento de F\'{\i}sica Te\'orica C-XI}}\vskip2pt
\centerline{{\it Universidad Aut\'onoma de Madrid}}\vskip2pt
\centerline{{\it 28049 Madrid, Spain}}

\vskip .7in

\baselineskip24pt
\indent

We use the lattice cooling method to investigate the
structure of some gauge fixed  $SU(2)$ Yang-Mills classical solutions of
the  euclidean equations of motion which are defined in the 3-torus
with symmetric twisted boundary conditions.

\end{titlepage}

\section{Introduction}
\setcounter{equation}{0}
\def\theequation{\thesection.\arabic{equation}}

\bigskip

In this paper we will analyse an $SU(2)$  Yang-Mills field configuration
which  is periodic in 3-space and tends to a pure gauge in
both $t=\pm \infty$. This configuration is a solution of the classical
euclidean equations of motion and has infinite action and topological
charge ( finite within one unit cell). The configuration
arises naturally when one studies
gauge fields on the spatial torus with twisted boundary conditions
(t.b.c.) [1].
 In particular, we fix the torus to have equal period in all three
directions and a twist vector $\vec{m} = (1,1,1)$. The presence of the
torus breaks down the $SO(3)$ rotational symmetry  down to the cubic
group. If we fix, without loss of generality, the length of the spatial
torus to $l=1$, we may write the boundary conditions

\begin{equation}
   {\bf A}_\mu (x + \hat{e}_i) = \sigma_i {\bf A}_\mu(x) \sigma_i
\end{equation}
where $\sigma_i$ are the Pauli matrices and $\hat{e}_i$ the unit vector
in the $i^{\underline{th}}$ direction. The choice of the Pauli matrices
as the twist matrices can be regarded as a partial gauge fixing of the
problem. The remaining group of gauge transformations must satisfy

\begin{equation}
{\bf \Omega}(x+\hat{e}_i) = \sigma_i {\bf \Omega}(x) \sigma_i
\end{equation}
This is not, however, the most general internal symmetry transformation
of our \linebreak  problem. One can perform a transformation

\begin{equation}
{\bf A}_\mu (x)\  \rightarrow \ {\bf A}'_\mu (x) = {\bf
\Omega}^{(z)}(x) {\bf A}_\mu(x) {\bf \Omega}^{(z) \dagger}(x) + i \
{\bf \Omega}^{(z)}(x)\partial_\mu {\bf \Omega}^{(z)\dagger}(x)
\end{equation}
with
\begin{equation}
{\bf \Omega}^{(z)}(x+\hat{e}_i) = z_i \ \sigma_i {\bf \Omega}^{(z)}
(x) \sigma_i \end{equation}
and $z_i= \pm1$. This transformation preserves the b.c. Eq.(1.1) and
leaves
the Yang-Mills action invariant. Nevertheless, if $z_i \neq 1$ for some
$i=1,2,3$ the transformation is not, strictly speaking, a gauge
transformation since it
modifies the values of the Wilson loops associated with non-contractible
loops (Polyakov loops):
\begin{equation}
 W(\gamma) \ \rightarrow \  W'(\gamma) =
z_i^{\omega_i(\gamma)}\  W(\gamma) \end{equation}
where $\omega_i(\gamma)$ is the winding number of the loop in the
$i^{\underline{th}}$ direction. We will call these transformations
singular gauge
transformations and the group of these transformations (modulo ordinary
gauge transformations) is isomorphic to $Z_2^3$. As  we will see later,
these transformations act non-trivially on our gauge field configuration
and thus we have 8 configurations rather than one.

 The configuration with smallest action (S=0) with our boundary
conditions is of the pure gauge form:
\begin{equation}
{\bf A}_\mu(x) = i\ {\bf\Omega}^{(z)}(x) \partial_\mu
{\bf\Omega}^{(z)\dagger}(x)
\end{equation}
with ${ \bf \Omega}^{(z)}(x)$ satisfying Eq.(1.4). However, it turns
out  [2]
 that there are only two gauge non equivalent configurations of this
type, those with $z_i=1$ for all $i$, and $z_i=-1$ for all $i$. In the
${\bf A_0}=0$ gauge these configurations are constant in time and equal
to  the minima
 of the potential energy $\sum_i TrB_i^2$. The problem becomes
that of a  $Z_2$ symmetric potential with two non-symmetric minima.
The order parameter distinguising these two is the Polyakov loop winding
the torus once along each direction  ($\omega_i=1$). Just as in the case
of the $\lambda \phi^4$ potential (and negative mass squared), we can
investigate the solution which interpolates between the two minima as
time goes from $-\infty$ to $+\infty$. This is an instanton solution and
gives rise to the leading weak coupling contribution to the energy
splitting   between the $Z_2$ symmetric and antisymmetric states.

 These instanton configurations are precisely the solutions which we are
 after. They have half integer topological charge and are precisely the
minimum
action configurations in a 4-torus with infinite length and with t.b.c.
both in time and space.
 The twist in time $n_{0i}=1$ eliminates the constant minima
solutions.

 The existence of these configurations was proved by Sedlacek [3].
In a
previous paper [4] we investigated the behaviour of these solutions
by using the lattice approximation and the cooling method [5].
There it
was shown that the total action and energy of the lattice minimum action
configurations scale towards a continuum value. The value of the
continuum
action is very close to $4\pi^2$, the absolute minimum for a
topological charge $Q$
of  $1/2$. Indeed, this minimum is only attained by a self-dual or
anti-self-dual configuration, so our configurations must be very
approximately,
if not exactly, self-dual. If we perform a parity transformation to this
self-dual instanton configuration we get an anti-self-dual anti-instanton
configuration with $Q=-1/2$.

 The purpose of this paper is to report a more extensive and accurate
analysis
of the lattice minimum configurations in order to answer some questions
which
were not settled by our previous paper. Apart from giving additional
support  to
the question of scaling towards the continuum and self-duality, we have
shown  that there are indeed eight different gauge-inequivalent
configurations (four
instantons and four anti-instantons) which can be obtained by acting
with the
group of singular gauge transformations on one of them. All these
configurations are non abelian and cubic-symmetric. We have obtained
functional expressions for the various physical quantities, including
the vector potentials and field strengths in a suitable gauge, which
describe
their qualitative features. Satisfactory quantitative description of
the data has been obtained by using these functional expressions and the
first few terms in a Fourier expansion of the functions involved. We
also report the steps that we have taken in the direction of finding an
analytic expression for the solution. Although our attempts in this
respect have not achieved the ultimate goal, we have
explored several paths involving
some lenghthy and non-trivial manipulations, and our results can be of
great help for future investigations.

Our solution, being non-abelian, is very different to the abelian
solutions [6] which are known to exist for other twists,
other values of the
topological charge and additional periodicity in time. However, it is
much more similar to the instanton solution on $S_{3} \times \Re$.\\
In a recent
paper [7] this solution has been used to investigate the
$\theta$-dependence beyond the steepest descent.
It is
in this spirit that our solution is physically relevant. The reader is
addresed   to Ref.7 and the earlier Ref.8 to see the role of our
instanton solution for the Yang-Mills dynamics, in the weak coupling
limit.

\section{ Analysis of the data}
\setcounter{equation}{0}
\def\theequation{\thesection.\arabic{equation}}

  Our strategy has been explained in Ref.4. We consider an $N_s^3\times
N_t$ lattice with twisted boundary conditions. We use  cooling   to
obtain configurations with smaller and smaller action. We stop once the
value of the action has changed by less than $10^{-4}$ over the last
100 cooling sweeps. To perform the analysis reported here we have used
the
configurations which we employed in Ref.4 together with
new data which extend over larger lattices $(N_s, N_t) = (11,29)$ and
 $(15,29)$. We have little to add to the results of Ref.4 concerning
integrated quantities. The new data just give extra support to the
conclusions presented there. The value of the action  $S$ is $39.234$
and $39.347$, for $(11,29)$ and $(15,29)$ in agreement with our
extrapolation  formula  $S=4 \pi^2 -29.7 a^2 $ (with $a=1/N_s$). The
degree of self duality is
now $X =2 \frac{|S_E-S_B|}{S_E+S_B} =0.0045 $    and $0.0033$ for the
same two lattices.
 We notice that the errors introduced by the lattice approximation are
smaller than 1\% for integral quantities. This can be explained by the
fact that they are $O(a^2)$. In the case of the vector potentials and
the
colour electric and  magnetic fields at a given point $A_i^a(x)$,
$E_i^a(x)$
and $B_i^a(x)$, we expect  errors of order a: typically then 10\%
for
our values of $N_s$. One can estimate more precisely the size of these
errors by several methods. For example comparing $E_i^a$ and $B_i^a$
at the same point the difference is never greater than 0.14 (
representing
3\% of the magnitude at those points). On the average a difference of
order 0.04 makes
the $\chi^2$ per degree of freedom of the comparison of order 1. Other
estimates  of the errors can be deduced by comparing values for
different $N_s$ or by using different lattice approximants to the
continuum value. The resulting errors depend on the quantity under
consideration
and on the value of t, but stay   within a factor of two of 0.04 in all
cases. We have chosen, thus, the value of 0.04 as a typical error which
we will be using in all $\chi^2$ fits of the data.

\subsection{Gauge invariant quantities }

  We start analysing the structure of our solution by looking at gauge
invariant quantities. Consider first the colour electric and magnetic
fields $E_i^{a}$, $B_{i}^{a}$. They can be regarded as six vectors
in the
3-dimensional colour space. Gauge transformations amount to rotations in
this space. Thus, the gauge invariant quantities are the scalar products
and moduli of these vectors $ M_{ij} = \vec{E}_i\cdot\vec{E}_j$. Due to
the approximate self-duality  the colour magnetic fields give   the same
information within errors. Before stating the results of our analysis
let us mention
one technical point. Our interest is centered on the values of these
quantities in the continuum limit. To extract these values one can use
various lattice quantities all of which differ by an amount of order
$a$. Nevertheless, in checking certain symmetries, it is essential
to choose  lattice observables which respect the lattice symmetry. For
that reason we have extracted the colour field strength    at each
lattice point, $F_{\mu \nu}^b (na)$, by averaging the four plaquettes
attached to that point in each  $\mu\nu$ plane:
\begin{equation}
     F_{\mu\nu}^{b} (na) = \frac{1}{4a^2}
                         Tr \{-i \sigma_{b} \ [{ \bf P}_{\mu\nu}(n)
 + { \bf P}_{\nu-\mu}(n) + { \bf P}_{-\nu\mu}(n) +  { \bf
P}_{-\mu-\nu}(n) ]\}
\end{equation}
where, for instance, ${\bf P}_{\nu-\mu}(n)$ stands for  the following
product of lattice links:
${\bf P}_{\nu-\mu}(n) = {\bf U}_{\nu}(n) {\bf U}_{\mu}^{\dagger} (n -
\hat{\mu}+\hat{\nu}) {\bf U}_{\nu}^{\dagger}(n-\hat{\mu}) {\bf U}_{\mu}
(n-\hat{\mu}) $.
 The previous quantity transforms in the right way under
gauge transformations and is symmetric under cubic transformations
around the point $na$.

   The main features of the solution are the following:
\begin{enumerate}
\item[a)]
At each time value there is a point of maximum energy density, which we
will call the spatial center of the instanton, whose coordinates are
time-independent and we will choose them as $\vec{x} = \vec{0}$.
Furthermore there is a time where the energy is maximal and we will
choose it as the origin of time, $t=0$. Our solution is cubic invariant
with respect to the spatial center and invariant under time reversal.
The spatial center is not one of the lattice points, but it can be
located by interpolation. The same is true for the origin of time.

\item[b)]
For all times and at the spatial center of the instanton one has $M_{ij}
\propto \delta_{ij}$. This orthogonality implies that the solution is
not abelian.  Away from the center the vectors $E_i^a$ cease to be
mutually orthogonal but they are never collinear.

\item[c)]
We have fitted the moduli and scalar products at $t=0$
  to the first terms in a Fourier expansion. Our results are

 \begin{eqnarray}
M_{11}&=&  \cos^2(\pi x)\ \ \{\ 15.00 + 20.02 \ \ [\phi(y) +
\phi(z)] \nonumber \\
  &   & -3.12 \ cos^2(\pi y) cos^2(\pi z) + 0.128\ sin^2(2\pi y)
sin^2(2\pi z) \nonumber  \\
  &   & -0.18\ [ cos^2(\pi y) sin^2(2\pi z) + cos^2(\pi z)
sin^2(2\pi y)]\} \\
  &   & + 32.74\ \ \phi(y) \phi(z)   \nonumber \\
  &   &  - sin^2(2\pi x)   \ \      \{\ 0.232 + 2.25 \ [cos^2(\pi
y)
+ cos^2(\pi z) ] \nonumber \\ &   &+ 0.80\  cos^2(\pi y)
cos^2(\pi z) \} \nonumber \\
\nonumber \\
M_{12} &=& - 1.608 \ [1-0.433 \ cos(2\pi z)] \ \chi(x)
\ \chi(y)\end{eqnarray}
where
   \[ \phi(x) = cos^2(\pi x) - 0.082 \ sin^2(2\pi x)\]
   \[ \chi(x) = sin(2\pi x) + 0.086 \ sin(4\pi x)\]

 The quality of the fit is very good ( see fig.1 for example). The
remaining components are just obtained by the apropiate replacement of
the indices and variables in the previous expression, in a way which is
consistent with cubic invariance and parity invariance.
Notice that $\vec{E}_i^2$ vanishes
when $x_i = \pm 1/2$ and $x_j = \pm 1/2$. Notice also that the solution
seems perfectly smooth.
\end{enumerate}

      We turn now our attention to non local gauge invariant
quantities:
 the Polyakov loops. Due to the boundary conditions these
variables are defined as

\begin{equation}
     {\bf \Omega}_\mu (x) = \ Texp \{i \int_{\gamma_{\mu}(x,x')} {\bf
A}_\nu dx^{\nu} \}
                    \  \  {\bf \Gamma}_\mu(x') \ Texp \{i
\int_{\gamma_{\mu}(x',x)} {\bf A}_\nu dx^{\nu} \}
\end{equation}
where $\gamma_\mu(a,b)$ is a straight line in the positive $\mu$
direction starting at $a$ and ending at $b$, $x'$ is the border of the
torus patch, $Texp$ is the ordered exponential and ${\bf \Gamma}_\mu(x)
$ the twist matrices. On the lattice these quantities are
simply
given by  the ordered product of the $\mu$-links corresponding to the
path $\gamma_\mu(x)$. These quantities transform like ${\bf E}_i$ under
gauge
transformations   and therefore the gauge invariant quantities are just
the scalar products in colour space $\frac{1}{2} Tr({\bf\Omega}_\mu (x)
{\bf \Omega}_\nu (x)) = X_{\mu\nu}(x)$ and the traces themselves
$\frac{1}{2} Tr({\bf \Omega}_\mu(x)) = X_\mu(x)$.

Since our instanton evolves in time from one pure gauge Eq.(1.6) ($z_i =
1$) to another gauge inequivalent one ($z_i = -1$), and a
representative of each
class is given by ${\bf \Omega}_i(x) = \pm i \sigma_i \ \ (i=1,2,3) $,
we
expect $X_{ij} = -\delta_{ij}$, $X_i = 0$ for large $|t|$. Indeed our
results agree with
this situation.
 As a matter of fact $X_\mu$ can be fitted with a $\chi^2=
0.005$ per degree of freedom ( with
an absolute error of 0.04) to
the formula
\begin{equation}
       X_\mu = \prod_{\nu\neq \mu} m_\nu \end{equation}
where $m_i= m(x_i)$ for $i=1,2,3$  and $m_0 = m_0(t)$. This is a very
remarkable factorization property. The function $m$ is parametrised as
\begin{equation} m(x) = cos(\pi x)( 1 + C sin^2(\pi x))\end{equation}
where $C=-0.196$
and $m_0(t)$ is given in Fig.2. Notice as well that the $m$ functions
are antiperiodic. Thus, by translating our solution by one period in any
direction we get a new solution. In this fashion one generates 4
solutions out of one. Thus, we are not dealing with a unique instanton
solution but rather with a family of solutions related by singular gauge
transformations.

\subsection{Gauge dependent quantities}

  In order  to extract all the information from our numerical results we
have to obtain also gauge dependent information. In particular, one
would like to extract the gauge potentials themselves in a suitable
gauge. This turns out to be feasible and our results are presented in
what follows.

 First of all one must introduce a simple gauge fixing,
which respects most of the symmetry of the problem. It is natural to
select ${\bf A}_0=0$, a condition which
 preserves all spatial symmetries.
To completely specify the gauge one has to fix the remaining
time-independent
gauge transformations. This can be done by fixing $\vec{A}(\vec{x},
t=-\infty)=\vec{0}$ or equivalently  ${\bf \Omega}_i (\vec{x},
t=-\infty)
=
i \sigma_i$. This fixes the gauge completely including global
gauge rotations.
At $t=+\infty$      the gauge fields must coincide with a pure gauge
Eq.(1.6) and in the $A_0=0$ gauge this gauge transformation is indeed
given by
the temporal Polyakov loop ${\bf \Omega}_0 (\vec{x},t=-\infty)$ .
Notice that ${\bf \Omega}_0$ transforms precisely like Eq.(1.4) for
$z_i=-1$.
 The exact form of ${\bf \Omega}_0(\vec{x},
t=-\infty)$ can be obtained in terms of the gauge invariant quantities
$\frac{1}{2}Tr({\bf \Omega}_0(\vec{x},t=-\infty))$ and
$\frac{1}{2}Tr({\bf\Omega}_0(\vec{x},t=-\infty)\cdot
{\bf \Omega}_i(\vec{x},t=-\infty))$.
These have been obtained from our data  and the result
can
be summarised by giving a functional form which describes  these data
with a $\chi^2=0.01$ per degree of freedom

\begin{equation}
      {\bf \Omega}_0 (\vec{x},t=-\infty) = m(x)m(y)m(z)\  I + i \
\sqrt{1-\prod_{i}
m_i^2 }\ \ \frac{\vec{f}\cdot\vec{\sigma}}{|\vec{f}|} \end{equation}
where $f_i= f(x_i)$ and $f(x) = sin(\pi x) (1+ B cos^2(\pi x))$, with
$B=-0.172$. Notice that $f(x)$ is very similar to $m(x+1/2)$. Changing
the value of $B$ to $C$ does not change significantly the quality of the
fit.

 Now we turn our interest towards computing ${\bf A}_i(\vec{x},t)$ in
the
previously mentioned gauge. To extract these quantities from our lattice
results implies some lengthy procedure. Apart from the typical order $a$
errors we have to deal with some other sources of errors.
Some errors arise through  the gauge fixing procedure due to the finite
value of $N_t/N_s$. In this case ${\bf A}_i$ is never a pure gauge and
cannot be set to zero. Nevertheless these errors are for most purposes
quite small and can be monitored by varying $N_t/N_s$. Our procedure to
evaluate the gauge fixing on the lattice configuration is as follows.
To  evaluate the gauge fixed value of a lattice link ${\bf U}_\mu(n)$,
we
choose a lattice path going through this link and starting and ending at
some point $P$. This point is situated on the surface of
minimum energy
( smallest time $n_t = - (N_t-1)/2$) and the path is made of three
pieces
$\gamma_1$, $\gamma_2$ and $\gamma_3$. The first and last paths are
situated on the surface of time coordinate $-(N_t-1)/2$.
$\gamma_1$ joins $P$ with $P'= (-\frac{(N_t-1)}{2}, \vec{n})$, the point
with equal spatial coordinates as the origin of the link ${\bf
U}_\mu(n)$.
$\gamma_3$ joins $P" = (-\frac{(N_t-1)}{2}, \vec{n}+ \hat{\mu})$ with
$P$. $\gamma_2$ moves from $P'$ along the positive time direction,
then goes through the link in question, and then back to $P"$ along
the negative time direction. We complete our gauge fixing by performing
a global gauge
transformation ${\bf \Omega}(P)$ which rotates the spatial Polyakov
loops ${\bf \Omega}_i$ starting at point $P$ to the values $i \sigma_i$.
Summarising
these transformations we write down the expressions for the gauge
transformed link ${\bf U}'_\mu(n)$:

\begin{equation} {\bf U}'_\mu(n) = {\bf \Omega}(P) {\bf U}_{\gamma_1}
{\bf U}_{\gamma_2} {\bf U}_{\gamma_3}
{\bf \Omega}^{\dagger}(P)\end{equation}
Indeed, if the spatial links at the smallest time are a pure gauge
configuration, ${\bf U}_{\gamma_1}$ and  ${\bf U}_{\gamma_3}$ can be
gauged to $I$.
And since in the temporal gauge ${\bf U}_{\gamma_2} = {\bf U}_\mu(n)$,
we see that ${\bf U}'_\mu(n)$ is really our gauge fixed link.

In addition, the position of $P$ and the choice of $\gamma_1$ and
$\gamma_3$ are irrelevant. However, as mentioned previously we do not
exactly have a pure spatial gauge for $-(N_t-1)/2$ and the actual choice
of  $\gamma_1$ and $\gamma_3$ is relevant. To preserve the fact that
${\bf U}'_\mu(n)$ (Eq.(2.8)) is a gauge transformed of ${\bf U}_\mu(n)$,
one must choose $P$, $\gamma_1$
and $\gamma_3$   in a predetermined way for all lattice
links. Our choice is as follows. The path $\gamma_1$ is obtained
by running first in the 1 direction, next in the 2 direction and
finally in the 3 direction. We finally obtain:

\begin{equation}    {\bf U}'_\mu(n) = {\bf \Omega}(P) {\bf L}(n)
{\bf U}_\mu(n) {\bf L}^{\dagger}(n+\hat{\mu})
{\bf \Omega}^{\dagger}(P)\end{equation}
with
\begin{eqnarray}    {\bf L}(n)& =& \prod_{q_1=0}^{n'_1-1}{\bf  U}_1(
P + q_1 \hat{e}_1) \cdot
\prod_{q_2=0}^{n'_2-1}{\bf U}_2( P+n'_1 \hat{e}_1 + q_2
\hat{e}_2) \cdot  \\
    & &     \prod_{q_3=0}^{n'_3-1} {\bf U}_3( P+ n'_1 \hat{e}_1 +
n'_2 \hat{e}_2 + q_3 \hat{e}_3) \cdot
                \prod_{q_0=0}^{n'_0-1} {\bf U}_0( P+ n'_1 \hat{e}_1 +
n'_2  \hat{e}_2 + n'_3 \hat{e}_3 + q_0 \hat{e}_0)  \nonumber
\end{eqnarray}
 where $n'_\mu = n_\mu - P_\mu $ and $P_\mu$ are the coordinates of
point $P$.

  Once the data are gauge fixed we can extract the lattice approximants
to the continuum fields. The potentials $A_i^a$ are taken to lie on the
mid-point of the corresponding links. The magnetic fields $B_i^a$,
are computed by the plaquette averages mentioned before. Our
results show the same qualitative features for both vector potentials
and magnetic fields. These features are described by the following
functional form which fits the data:

\begin{eqnarray}
 O_1^1  & = & q_1(t) \ \cos(\pi y) \ \cos(\pi z) \  ( 1 + q_2(t)
cos(2\pi x))
        \nonumber     \\
        &   &( 1 + q_3(t)\  cos(2\pi y))\ (1 + q_3(t) cos(2\pi z))
         \nonumber\\
        \nonumber \\
 O_1^2  & = & ( 1 + a_1(t)\ cos(2\pi x))\   (1 + a_2(t) cos(2\pi z))
        \nonumber         \\
        &   &  [ \ a_3(t)\ sin(\pi z)\ cos(\pi x)\ (1 + a_4(t) cos(\pi
y))
                  \\
        &   &    + a_5(t)\ cos(\pi z)\ sin(\pi x)\ sin(2\pi y) ]
        \nonumber\\
        \nonumber \\
 O_1^3  & = & ( 1 + a_1(t)\ cos(2\pi x))\   (1 + a_2(t) cos(2\pi y))
       \nonumber  \\
        &   &  [ -a_3(t)\ sin(\pi y)\ cos(\pi x) \ (1 + a_4(t) cos(\pi
z))
         \nonumber        \\
        &   &   + \ a_5(t)\ cos(\pi y) \ sin(\pi x) \ sin(2\pi z) ]
\nonumber
 \end{eqnarray}
where $O_i^a$ stands for either $B_i^a$, $E_i^a$ or $A_i^a$, but
with different values of the parameters $q_i$ and $a_i$. The remaining
components of $O_i^a$ can be obtained by cyclic permutations of the
indices 1,2,3 in Eq.(2.11). The values of the parameters appearing in
the expression are given in tables 1 to 4.

The previous functional form describes
satisfactorily the data with a maximum $\chi^2$ per degree of freedom
equal to 4 for $A_i^a$ and 6 for $B_i^a$.
 It can be used to study the main
features of the solution.  Another interesting aspect of the formula is
to  serve as a guide for the construction of an ansatz which could lead to
an
analytic expression of the solution. In the following section we will
describe the steps that we have taken in this direction.

\section{Ansatz}
\setcounter{equation}{0}
\def\theequation{\thesection.\arabic{equation}}

   In Eq.(2.11) we have given an analytic expression which describes our
numerical results within errors.  It is clear that the solution is not
as simple as that expression, but we want to explore the consequence of
the fact that it shares with it the same behaviour under the symmetry
operations of the system.  Thus, we will assume the following general
form for the solution (for all t):

\begin{equation}
A_i^a= \delta_i^a Q^{(1)} (t,x_i,x_j,x_k) +
       (\epsilon^{iab})^2 Q^{(2)} (t,x_i,x_a,x_b) +
       \epsilon^{iab} Q^{(3)} (t,x_i,x_a,x_b)
\end{equation}
where $Q^{(1)}(t,x,y,z)=Q^{(1+)}(t,x,y,z)$ is symmetric under the
exchange of $y$ and $z$,
even in all the three variables, antiperiodic in $y$ and $z$, periodic
in $x$. $Q^{(2)}
(t,x,y,z)$ is odd in $x$,$y$, even in $z$, antiperiodic in $x$,$z$ and
periodic in $y$. $Q^{(3)} (t,x,y,z)$ is odd in the third variable and
even in the other two, antiperiodic in $x$,$z$ and periodic in $y$. The
previous expression implies that the solution is cubic symmetric with
$A_i^a(\vec{x})$ and $B_i^a(\vec{x})$ transforming in the $T_1\otimes
T_1$ representation of this group ($T_1$ is just the spin 1
representation of rotations). In addition, the behaviour under
translations by one period is just what is required by the twisted
boundary conditions.
      The previous properties are more restrictive than what can be
deduced by these symmetry properties alone. More precisely
$Q^{(1)}(t,x,y,z)$ could contain an additional term $Q^{(1-)}(t,x,y,z)
$ which is odd in
all three variables, antiperiodic and antisymmetric   in the two last
variables $y$,$z$. To test the presence or absence of this term we have
added to $A_1^1$ in the parametrization Eq.(2.11) a term

\begin{equation}
q_4 \ sin(2\pi x ) sin(\pi y) sin(\pi z)\  (cos(2\pi y)- cos(2\pi z))
\end{equation}
and refitted our data for ${\bf A}$ and ${\bf B}$. The
presence of a
such a term improves the value of the $\chi^2$ by a factor $2/5$, but
the size of
$q_4$  and thus its contribution is of the order of the estimated size
of the $O(a)$ errors. Thus, we will keep an open attitude and explore
both possibilities.

  Another additional information which might be crucial for finding the
analytic expression is the behaviour  under $P \cdot T$. Both parity
and time
reversal map instantons into anti-instantons, but one could ask whether
the product of these two leaves our solution invariant or not. The
implications of this invariance are complicated by the future-past
asymmetry of our gauge choice. Invariance under $P\cdot T$ implies

\begin{equation}
-{\bf A}_i(-\vec{x},-t) = {\bf \Omega}_0^\dagger
{\bf A}_i(\vec{x},t)  {\bf \Omega}_0 + i {\bf \Omega}_0^\dagger
\partial_i  {\bf \Omega}_0
\end{equation}
where ${\bf \Omega}_0$ is the gauge matrix which describes the pure
gauge at $t= \infty$. The matrix  ${\bf \Omega}_0$   was parametrised
by Eq.(2.7) and indeed the form of the axis of rotation
$\vec{f}/|\vec{f}|$ is precisely consistent with the restricted form
where $Q^{(1)}= Q^{(1+)}$  even in all  three variables. The
necessary
and sufficient condition for the gauge field at $\infty$ to be of the
restricted form $Q^{(1-)}=0$ is that $f_i$ is a function of $x_i$ alone.

 Expression (3.3) conflicts nevertheless with the requirement
$Q^{(1-)}=0$   since in principle a rotation by ${\bf \Omega}_0$ does
not preserve this constraint. If we impose that a general rotation
around $\vec{f}/|\vec{f}|$ should preserve $Q^{(1-)}=0$ we must have

\begin{eqnarray}
Q^{(2)} (t,x,y,z)& = & f(y) \hat{Q}^{(2)}(t,x,y,z) \\
Q^{(3)} (t,x,y,z)& = & f(z) \hat{Q}^{(3)}(t,x,y,z)
\end{eqnarray}
where $\hat{Q}^{(2)}(t,x,y,z)$ and  $\hat{Q}^{(3)}(t,x,y,z)$  are
symmetric
functions of the last two arguments  $y$,$z$. We have verified that
these additional constraints are reasonably well satisfied by our
data. Thus, we arrive at  a restricted parametrization
consistent with our data:

\begin{equation}
A_i^a (\vec{x},t) = \rho_1^i\  \frac{f_i f_a}{|\vec{f}|} + \rho_2^i
\ |\vec{f}| ( \delta_i^a - \frac{f_i f_a}{|\vec{f}|^2}) +
\rho_3^i \  \epsilon_{abi} f_b
\end{equation}
where  $\rho_1^i$, $\rho_2^i$ and $\rho_3^i$ must be  even
functions of the
three variables $x$,$y$,$z$. $\rho_1^i$ and $\rho_2^i$  are periodic
in $x_i$ and antiperiodic in $x_j \ne x_i$, and $\rho_3^i$ is
antiperiodic in $x_i$ and periodic in $x_j \ne x_i$.

The fact that this form is not a mere consequence of the symmetry
properties implies that, when requiring self-duality of such a solution,
new equations will arise which guarantee the form to be preserved.

    To start with, let us perform a time independent gauge
transformation and write

\begin{equation}
{\bf A}_i'(\vec{x},t) = {\bf \Omega}_0^{-1/2} {\bf A}_i
(\vec{x},t) {\bf
\Omega}_0^{1/2} + i {\bf \Omega}_0^{-1/2}\partial_i {\bf
\Omega}_0^4{1/2} \end{equation}
This new vector potential  complies also  to the form Eq.(3.6) in what
respects to the
behaviour of the $\rho'$ functions under parity.
In addition
it must satisfy $ -{\bf  A}'_i(-\vec{x},-t) = {\bf A}'_i (\vec{x},t)$.
This implies that $\rho_1^{'i}$ and $\rho_2^{'i}$  are odd in t and
$\rho_3^{'i}$ is even in t. We have explicitly checked these properties
in our nummerical data and found agreement within errors. In fact, given
self-duality it is enough that $\rho_1^{'i}=\rho_2^{'i}= 0$  at
$t=0$ to
guarantee the appropiate behaviour for $t\ne 0$. At $t=0$ we have been
able to fit both ${\bf A}_i$ and ${\bf B}_i$ with $\rho_3^{'i}$ alone.

 Imposing now self-duality we obtain six equations rather than three.
The first set of equations follows from the requirement that the
$\rho_a^{'i}$ should have the required behaviour under parity. The
equations are:

\begin{equation}
\epsilon^{ijk} (\nabla_j \hat{\rho}_a^{'k}  + T^{abc}
\hat{\rho}_b^{'j} \hat{\rho}_c^{'k})=0
\end{equation}
where $\nabla_j = \frac{1}{f_j} \partial_j$, $\hat{\rho}_a^{'i} =
\rho_a^{'i}+ \delta_{a3} \partial_i(f(x_i))/(2|\vec{f}|^2)$   and
$T^{abc}$ is a
completely antisymmetric tensor with $-T^{123} = T^{231} =
T^{312}= 1$.

The solution to this equation is as follows
\begin{equation}
\hat{\rho}_a^{'i} = R^{a\rho \sigma} q_{\rho} \nabla_i q_{\sigma}
\end{equation}
where $q_{\rho}$ satisfy $ q_1^2+q_2^2-q_3^2-q_4^2 =1$ and
$R^{a \rho \sigma}$
is a tensor, antisymmetric in the last two  indices, with
$R^{121}=R^{134}=R^{213}=R^{242}=R^{314}=R^{323}=1$. The
functions $q_{\mu}$ then establish a mapping from $S_1^3 \times \Re $
 onto that hyperboloid.

  If we now plug the solution for $\hat{\rho'}$ Eq.(3.9)  into the
self-duality equation, and after considerable massaging, we arrive at

\begin{eqnarray}
\partial_0 s^a & = &  \epsilon^{abc}  s_b s_c' \\
\nabla_i s_a'& = & \nabla_j \nabla_k s_a  \ \ ( \forall i \ne j \ne k)
\\ s^a s_a & = & |\vec{f}|^2
\end{eqnarray}
where indices are raised and lowered with the metric $ g_{ab} = diag
(1,-1,-1)$  and $ s_1= |\vec{f}| (q_1^2+ q_2^2+ q_3^2+q_4^2)$,
$ s_2 = 2 |\vec{f}| (q_2 q_3 - q_1 q_4)$ and $s_3 = 2 |\vec{f}| (q_1 q_3
+ q_2 q_4)$. The dynamics Eq.(3.10) is consistent with the constraint
Eq.(3.12). The most important restriction follows from the
integrability
condition of Eq.(3.11), which together    with Eq.(3.12) serves to
restrict
the possible value of $f(x)$. If we introduce the function $z_i \equiv
z(x_i)$ by the condition $ \frac{d z(x)}{dx} = f(x)$, Eq.(3.11) forces
$s_a$ to be sum of single variable functions  of the variables   $a_1
z_1 + a_2
z_2 + a_3 z_3$ with $a_i = +1,-1$ or 0 and $\sum a_i = 1(mod 2)$. On the
other
hand $|\vec{f}|^2$ is a sum $ F(z_1) + F(z_2) + F(z_3)$. One class of
solutions is given by  functions $s_a( z_1+z_2+z_3)$. In this case
Eq.(3.12) forces $f(x) = A x$ and then the whole problem possesses
spherical
symmetry $z_1+ z_2+ z_3 \propto |\vec{x}|^2$. We are then in the
situation discused in Ref.9 and our equations recover all the axially
symmetric solutions including the B.P.S.T.[10] one. It is not easy to
find
other solutions of Eq.(3.11) and Eq.(3.12). Indeed, if we impose these
conditions on the coefficients of the Taylor expansion of the functions
$s_a(z)$, the number of conditions grows faster than the number of
parameters. This suggests to try polynomials for the functions $s_a$. In
this way we have discovered other solutions, most remarkably one which
is defined on the torus with $z=Acos(Bx+C)$ and where $s_1$ and $s_3$
are  polynomials
of second degree in $z_i$ and $s_2$ of first degree. The solution goes
to a pure gauge in $t\rightarrow \pm \infty$ but is singular at $t=0$
and the total action is divergent ( in the unit cell ). We have been
unable to find any solution which is satisfactory and coincides with our
numerical data.

      If our solution does not follow from Eqs.(3.10)-(3.12), one has to
drop some of the assumptions leading to those equations. Most likely it
is Eq.(3.6) with the assumed properties for $\rho^i_a$ which is wrong,
since its validity is simply based on the ability to describe the data.
Unfortunately, if we go back to the form Eq.(3.1) with $Q^{(1)} =
Q^{(1+)} + Q^{(1-)}$ we have no additional equations which could
allow
an analytic  development of the sort leading to Eqs.(3.10)-(3.12). One
can simply rewrite the self duality equations for the $Q$ functions, but
we do not know how to select solutions going to a pure gauge in $t=\pm
\infty$.

\section{Summary and conclusions}

  In this paper, we have studied the instanton-like solution which
occurs in a 3-torus with symmetric twist. It is remarkable that the
cooling method provides a very accurate description of this solution
which allows a systematic analysis of its properties. Our results imply
that there are indeed four instanton  and four anti-instanton solutions,
all of which are (within errors) cubic symmetric and $P\cdot T$
invariant. The solution is also smooth, although singularities in
sufficiently high derivatives are not excluded. We have given functional
forms which incorporate the previously mentioned properties, plus
additional restrictions consistent with the data. Nonetheless, the more
restrictive form, which allows the reduction of the problem to motion on
a hyperboloid and recovers many multi-instanton solutions on  $S_4$
does not seem to contain the solution of our problem.

\section*{Acknowledgements}

This work has been partially financed by CICYT.

\newpage

\section*{Table Captions}
\begin{enumerate}
\item [Table 1:]
The value of the parameters  obtained when fitting the functional form
(2.11) and (3.2) to the $N_s = 11$, $N_t = 29$ data for $A_1^1$.
$n_t=15$ corresponds to the time of maximum energy. Errors are indicated
by giving within parenthesis their magnitude affecting the last quoted
digit.
\item [Table 2:]
The same as Table 1 but for the parameters entering in $A_1^2$,
$A_1^3$. \item [Table 3:]
The same as Table 1 but for the data of $B_1^1$.
\item [Table 4:]
The same as Table 2 but for the data of $B_1^2$, $B_1^3$.

\end{enumerate}

\newpage

\section*{Figure Captions}
\begin{enumerate}
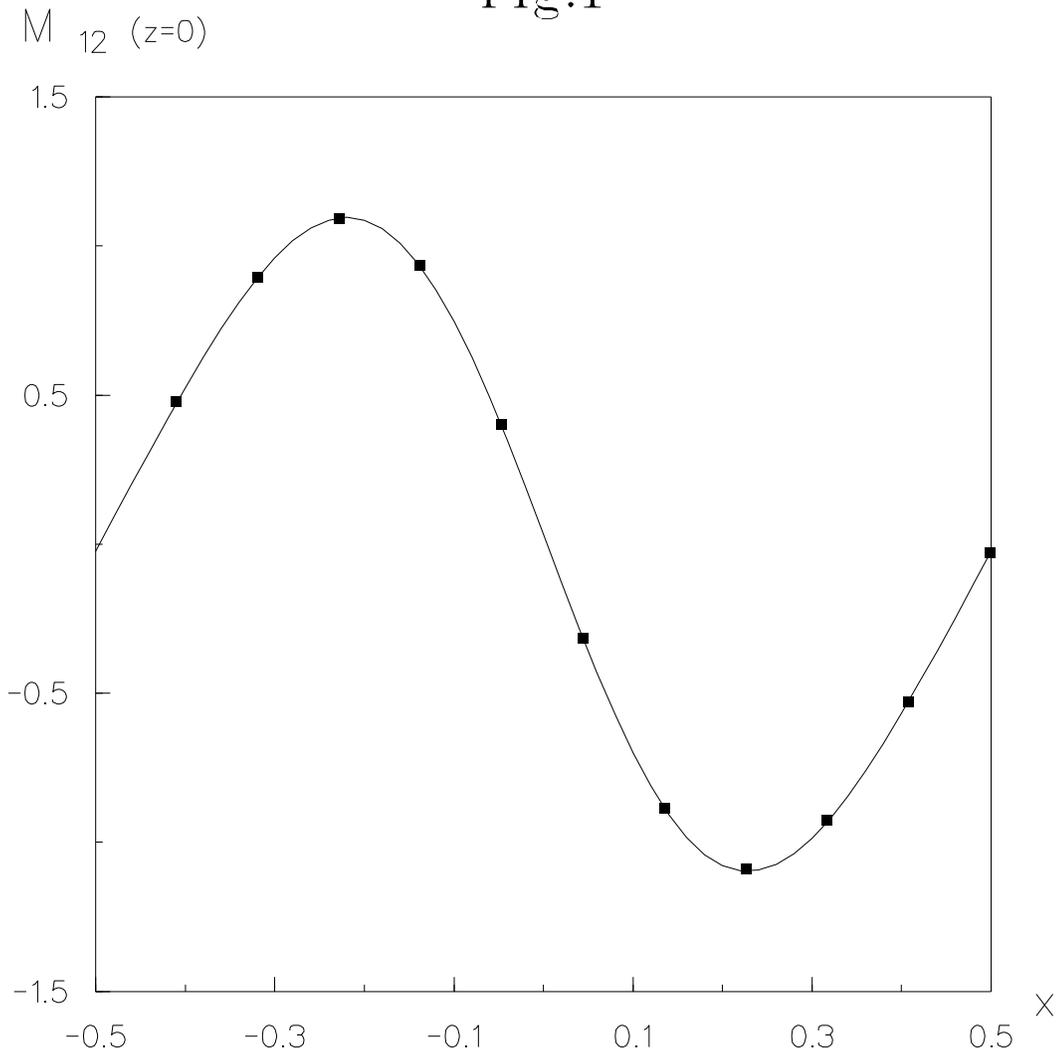
\item[Fig.1:]
The scalar product $M_{12}=\vec{E}_1\cdot \vec{E}_2$  plotted as
a function of $x$
for $y=0.209$ and $z=0$. The squares correspond to the lattice values
for
$N_s=11$, $N_t=21$. They are compared to the prediction of
Eq.(2.3).

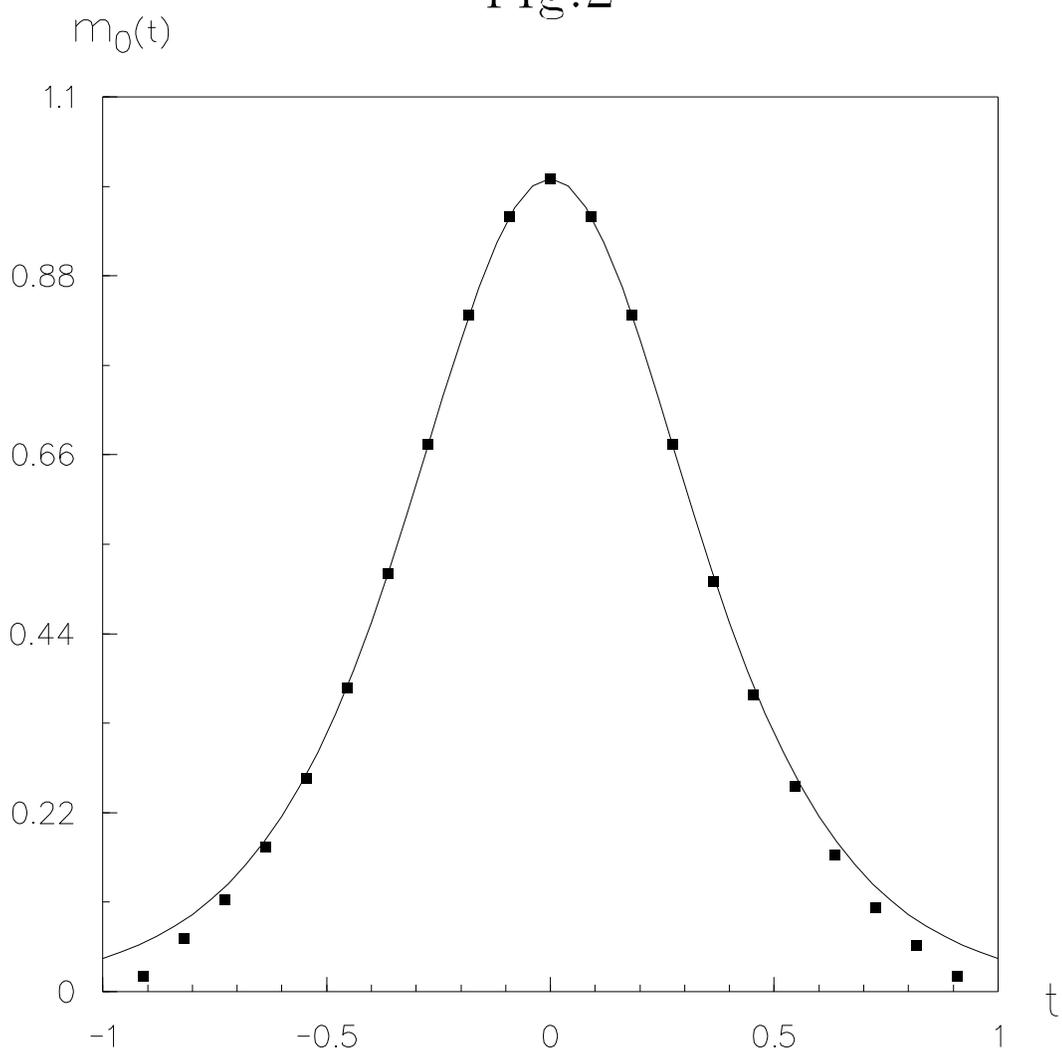
\item[Fig.2:]
$m_0(t)$ appearing in Eq(2.5) is plotted as a function of time for
$N_s=11$, $N_t=21$. The continous line represents the function
$cos(\frac{\pi}{2} tanh(0.69\pi t))$.

\end{enumerate}

\newpage

\begin{center}
\begin{tabular}{|c|c|c|c|c|}
\multicolumn{5}{c}{\bf {\huge Table 1}}    \\
  \hline \hline
{\bf nt} &${\bf  q_1}$& ${\bf q_2}$&$ {\bf q_3}$&$ {\bf q_4}$
 \\ \hline \hline
5    & 0.094(8) & 0.04(6)  & 0.02(6)  & 0.00(1)    \\   \hline
6    & 0.140(8) & 0.05(2)  & 0.02(2)  & 0.00(1)    \\   \hline
7    & 0.208(8) & 0.06(3)  & 0.02(3)  & 0.00(1)    \\   \hline
8    & 0.302(8) & 0.07(2)  & 0.02(2)  & 0.00(1)    \\   \hline
9    & 0.438(8) & 0.09(1)  & 0.03(2)  & 0.00(1)    \\   \hline
10   & 0.624(8) & 0.103(9) & 0.043(9) & 0.00(1)    \\   \hline
11   & 0.870(8) & 0.120(6) & 0.058(7) & 0.01(1)    \\   \hline
12   & 1.182(6) & 0.138(4) & 0.078(5) & 0.02(1)    \\   \hline
13   & 1.550(6) & 0.154(3) & 0.102(4) & 0.02(1)    \\   \hline
14   & 1.954(6) & 0.167(3) & 0.129(3) & 0.03(1)    \\   \hline
15   & 2.356(6) & 0.177(1) & 0.156(2) & 0.05(1)    \\   \hline
16   & 2.716(6) & 0.182(2) & 0.182(2) & 0.07(1)    \\   \hline
17   & 3.010(6) & 0.184(1) & 0.205(2) & 0.09(1)    \\   \hline
18   & 3.234(6) & 0.183(1) & 0.224(2) & 0.11(1)    \\   \hline
19   & 3.392(6) & 0.181(1) & 0.240(2) & 0.12(1)    \\   \hline
20   & 3.498(6) & 0.179(1) & 0.252(2) & 0.13(1)    \\   \hline
21   & 3.566(6) & 0.178(1) & 0.261(2) & 0.14(1)    \\   \hline
22   & 3.610(6) & 0.176(1) & 0.267(2) & 0.14(1)    \\   \hline
23   & 3.638(6) & 0.175(1) & 0.272(2) & 0.14(1)    \\   \hline
24   & 3.654(6) & 0.174(1) & 0.276(2) & 0.14(1)    \\   \hline
25   & 3.664(6) & 0.174(1) & 0.278(2) & 0.14(1)    \\   \hline
26   & 3.672(6) & 0.173(1) & 0.280(2) & 0.14(1)    \\   \hline
27   & 3.674(6) & 0.173(1) & 0.282(2) & 0.14(1)    \\   \hline
29   & 3.674(6) & 0.172(1) & 0.284(2) & 0.14(1)    \\   \hline
\end{tabular}
\end{center}

\newpage

\begin{center}
\begin{tabular}{|c|c|c|c|c|c|}
\multicolumn{6}{c}{\bf \huge  Table 2}    \\
  \hline \hline
{\bf nt} &  $ {\bf a_1}$& $ {\bf a_2}$&$ {\bf a_3}$&$ {\bf a_4}$&
${\bf a_5}$ \\ \hline \hline
6    & 0.00(6)  &-0.02(6)  & -0.122(6) & 0.07(4)  &-0.002(6) \\ \hline
7    & 0.01(5)  &-0.03(4)  & -0.172(6) & 0.09(3)  & 0.008(6) \\ \hline
8    & 0.01(3)  & 0.01(3)  & -0.244(6) & 0.11(2)  & 0.018(6) \\ \hline
9    & 0.01(2)  & 0.03(2)  & -0.348(6) & 0.14(2)  & 0.034(6) \\ \hline
10   & 0.02(1)  & 0.04(1)  & -0.496(6) & 0.15(1)  & 0.062(6) \\ \hline
11   & 0.02(8)  & 0.07(1)  & -0.704(6) & 0.19(8)  & 0.109(6) \\ \hline
12   & 0.033(6) & 0.092(5) & -0.984(6) & 0.225(6) & 0.174(6) \\ \hline
13   & 0.042(5) & 0.118(5) & -1.344(6) & 0.253(4) & 0.272(6) \\ \hline
14   & 0.052(4) & 0.145(4) & -1.776(6) & 0.278(3) & 0.402(6) \\ \hline
15   & 0.060(3) & 0.169(3) & -2.258(6) & 0.298(3) & 0.558(6) \\ \hline
16   & 0.065(3) & 0.189(2) & -2.746(6) & 0.311(2) & 0.730(6) \\ \hline
17   & 0.068(2) & 0.204(2) & -3.200(6) & 0.314(2) & 0.902(6) \\ \hline
18   & 0.069(2) & 0.215(2) & -3.590(6) & 0.324(2) & 1.056(6) \\ \hline
19   & 0.069(2) & 0.221(2) & -3.902(6) & 0.325(2) & 1.186(6) \\ \hline
20   & 0.067(2) & 0.225(1) & -4.138(6) & 0.325(1) & 1.290(6) \\ \hline
21   & 0.065(2) & 0.227(2) & -4.312(6) & 0.324(1) & 1.368(6) \\ \hline
22   & 0.064(2) & 0.229(1) & -4.434(6) & 0.323(1) & 1.424(6) \\ \hline
23   & 0.063(2) & 0.230(1) & -4.520(6) & 0.322(1) & 1.436(6) \\ \hline
24   & 0.062(2) & 0.230(1) & -4.580(6) & 0.321(1) & 1.494(6) \\ \hline
25   & 0.061(2) & 0.231(1) & -4.620(6) & 0.320(1) & 1.514(6) \\ \hline
26   & 0.061(2) & 0.231(1) & -4.648(6) & 0.320(1) & 1.530(6) \\ \hline
27   & 0.061(2) & 0.231(1) & -4.666(6) & 0.320(1) & 1.540(6) \\ \hline
29   & 0.061(2) & 0.232(1) & -4.686(6) & 0.320(1) & 1.554(6) \\ \hline
\end{tabular}
\end{center}
\newpage

\begin{center}
\begin{tabular}{|c|c|c|c|c|}
\multicolumn{5}{c}{\bf \huge  Table 3}    \\
  \hline \hline
{\bf nt} &${\bf q_1}$& $ {\bf q_2}$&$ {\bf q_3}$&$ {\bf q_4}$
 \\ \hline \hline
3    & 0.216(8) & 0.04(3)  & 0.00(3)  & 0.00(1) \\   \hline
5    & 0.424(8) & 0.06(1)  & 0.01(2)  & 0.00(1) \\   \hline
6    & 0.610(8) & 0.07(8)  & 0.02(1)  & 0.00(1) \\   \hline
7    & 0.876(8) & 0.090(5) & 0.026(8) & 0.00(1) \\   \hline
8    & 1.246(8) & 0.109(4) & 0.038(6) & 0.01(1) \\   \hline
9    & 1.736(8) & 0.130(3) & 0.054(4) & 0.01(1) \\   \hline
10   & 2.352(8) & 0.152(2) & 0.077(3) & 0.03(1) \\   \hline
11   & 3.058(8) & 0.175(2) & 0.107(2) & 0.05(1) \\   \hline
12   & 3.760(8) & 0.196(1) & 0.144(2) & 0.08(1) \\   \hline
13   & 4.300(8) & 0.213(1) & 0.191(2) & 0.12(1) \\   \hline
14   & 4.504(8) & 0.223(1) & 0.244(2) & 0.17(1) \\   \hline
15   & 4.278(6) & 0.224(1) & 0.301(2) & 0.21(1) \\   \hline
16   & 3.682(8) & 0.216(1) & 0.361(2) & 0.23(1) \\   \hline
17   & 2.900(8) & 0.200(2) & 0.421(2) & 0.21(1) \\   \hline
18   & 2.124(8) & 0.181(2) & 0.482(3) & 0.17(1) \\   \hline
19   & 1.474(6) & 0.160(3) & 0.542(5) & 0.12(1) \\   \hline
20   & 0.986(6) & 0.141(4) & 0.603(7) & 0.08(1) \\   \hline
21   & 0.644(6) & 0.124(6) & 0.663(1) & 0.04(1) \\   \hline
22   & 0.412(6) & 0.108(8) & 0.73(1)  & 0.03(1) \\   \hline
23   & 0.264(6) & 0.09(1)  & 0.79(3)  & 0.01(1) \\   \hline
24   & 0.166(6) & 0.08(2)  & 0.87(4)  & 0.00(1) \\   \hline
26   & 0.062(6) & 0.03(4)  & 1.1(1)   & 0.00(1) \\   \hline
\end{tabular}
\end{center}
\newpage

\begin{center}
\begin{tabular}{|c|c|c|c|c|c|}
\multicolumn{6}{c}{\bf \huge Table 4}    \\
  \hline \hline
{\bf nt} &  $ {\bf a_1}$& $ {\bf a_2}$&$ {\bf a_3}$&$ {\bf a_4}$&
${\bf a_5}$ \\ \hline \hline
3    & 0.00(6)  & 0.00(6)  & -0.120(6) & 0.07(4)  & 0.006(6) \\ \hline
5    & 0.01(2)  & 0.02(3)  & -0.292(6) & 0.10(2)  & 0.022(6) \\ \hline
6    & 0.01(2)  & 0.02(2)  & -0.438(6) & 0.12(1)  & 0.042(6) \\ \hline
7    & 0.02(1)  & 0.04(1)  & -0.644(4) & 0.147(8) & 0.074(6) \\ \hline
8    & 0.019(2) & 0.050(8) & -0.940(6) & 0.177(6) & 0.132(6) \\ \hline
9    & 0.027(5) & 0.069(5) & -1.354(6) & 0.210(4) & 0.224(6) \\ \hline
10   & 0.037(4) & 0.095(4) & -1.912(3) & 0.245(3) & 0.370(6) \\ \hline
11   & 0.050(3) & 0.126(3) & -2.626(6) & 0.280(2) & 0.584(6) \\ \hline
12   & 0.064(2) & 0.161(2) & -3.464(6) & 0.313(2) & 0.872(6) \\ \hline
13   & 0.079(2) & 0.198(1) & -4.322(6) & 0.341(1) & 1.212(6) \\ \hline
14   & 0.092(2) & 0.233(1) & -5.018(6) & 0.361(1) & 1.550(6) \\ \hline
15   & 0.100(2) & 0.262(1) & -5.354(6) & 0.372(1) & 1.796(6) \\ \hline
16   & 0.100(2) & 0.282(1) & -5.214(6) & 0.372(1) & 1.880(6) \\ \hline
17   & 0.093(2) & 0.293(1) & -4.754(6) & 0.364(1) & 1.786(6) \\ \hline
18   & 0.080(2) & 0.297(2) & -3.950(6) & 0.349(2) & 1.556(6) \\ \hline
19   & 0.064(3) & 0.294(2) & -2.998(6) & 0.331(2) & 1.264(6) \\ \hline
20   & 0.047(4) & 0.289(3) & -2.232(6) & 0.312(3) & 0.974(6) \\ \hline
21   & 0.031(5) & 0.282(4) & -1.608(6) & 0.293(4) & 0.724(6) \\ \hline
22   & 0.017(7) & 0.276(6) & -1.132(6) & 0.277(3) & 0.522(6) \\ \hline
23   & 0.00(1)  & 0.271(8) & -0.784(6) & 0.264(8) & 0.370(6) \\ \hline
24   & 0.00(1)  & 0.27(1)  & -0.536(3) & 0.26(1)  & 0.230(6) \\ \hline
25   & 0.00(2)  & 0.28(1)  & -0.364(6) & 0.26(2)  & 0.180(6) \\ \hline
26   & 0.00(3)  & 0.30(2)  & -0.244(6) & 0.27(3)  & 0.130(6) \\ \hline
29   & 0.00(3)  & 0.30(2)  & -0.244(6) & 0.27(3)  & 0.130(6) \\ \hline
\end{tabular}
\end{center}
\newpage



\end{document}